\documentclass[%
 reprint,
 amsmath,amssymb,
 aps,
prb,
]{revtex4-2}

\usepackage{graphicx}
\usepackage{dcolumn}
\usepackage{bm}
\usepackage{braket}
\usepackage{xcolor}

\newcommand\norm[1]{\left\lVert#1\right\rVert}

\begin{document}

\title{Topologically protected photovoltaics in Bi nanoribbons}

\author{Alejandro Jos\'e Ur\'ia-\'Alvarez}
\email{alejandro.uria@uam.es}
\affiliation{Departamento de F\'isica de la Materia Condensada, Condensed Matter Physics Center (IFIMAC), and Instituto Nicol\'as Cabrera (INC),
Universidad Aut\'onoma de Madrid, Cantoblanco 28049, Spain}

\author{Juan Jos\'e Palacios}
\affiliation{Departamento de F\'isica de la Materia Condensada, Condensed Matter Physics Center (IFIMAC), and Instituto Nicol\'as Cabrera (INC),
Universidad Aut\'onoma de Madrid, Cantoblanco 28049, Spain}

\date{\today}

\begin{abstract}
Photovoltaic efficiency in solar cells is hindered by many unwanted effects. Radiative channels (emission of photons) sometimes mediated by non-radiative ones (emission of phonons) are principally responsible for the decrease in exciton population before charge separation can take place. Another unwanted effect is electron-hole recombination at surfaces where in-gap edge states serve as the non-radiative channels. Topological insulators (TIs), in particular the most common ones characterized by a $\mathbb{Z}_2$ invariant, are rarely explored from an optoelectronics standpoint, possibly because of their typically small gaps. Otherwise, they are not much different from small-gap semiconductors and excitons are also expected to be their simplest excitations. Here we show that one can take advantage of the non-radiative decay
channel due to the surface states to generate a topologically protected photovoltaic current. Focusing on two-dimensional TIs, and specifically for illustration purposes on a Bi(111) monolayer, we show the potential of TI nanoribbons to generate edge charge accumulation and edge currents under illumination.
\end{abstract}

\maketitle


The creation of pairs of a free electron and a hole  may suffice to broadly explain the  optical conductivity of insulators and semiconductors, but, in general, bound electron-hole (e-h) pairs or excitons also play a non-negligible role \cite{louie_electron_hole, many_body_electronic, louie}. This is particularly true in two-dimensional (2D) crystals, where excitons become tightly bound due to the strong confinement and low screening. These include, for instance, hexagonal boron nitride \cite{excitons_hbn, exciton_hybrid_hbn} or transition metal dichalcogenides, which have been extensively studied in this regard \cite{RevModPhys.90.021001, mos2_macdonald, mos2_ridolfi}. Additionally, it has also been shown that non-linear phenomena such as high-harmonic generation \cite{high_harmonic} and bulk photovoltaic effects \cite{circular_photogalvanic_weyl,sipe} can be greatly enhanced in 2D crystals \cite{ges_excitons, huang2023, ruan2023excitonic, chang2023diagrammatic}. This, together with the high tunability of the atomic structure through strain \cite{quenching_esteve} and the ability to select excitations based on the light polarization render 2D optoelectronics as a very active field from both fundamental and technological perspectives \cite{Mueller2018}. 

Of particular interest is light-energy conversion in the form of photocurrent generation, on which solar cells devices are based.
The formation of bound e-h pairs and their subsequent separation is the most common source of photocurrent generation. In conventional solar cells, based on p-n junctions, this is achieved with the built-in electric field in the depletion zone, which separates the free charge carriers generating a chemical potential difference or a current depending on the circuit scheme. Since the efficiency of conventional cells is constrained by the Shockley–Queisser limit \cite{shockley},  alternative dissociation mechanisms have been proposed as in multijunctions cells \cite{multijunction_cells} or in excitonic solar cells, where a bound electron-hole pair is formed and diffuses to an interface where the charge separation takes place \cite{Gregg2003, multiexciton}.

Topological insulators (TIs), on the other hand, have garnered significant attention in recent years due to their potential for use in spintronic devices, among other more fundamental reasons \cite{spintronics}.
However, there has been relatively little study of TIs from an optical perspective \cite{exp_bi_conductivity, Bi2Se3_exciton_exp, Pandey2021, optical_abs_ti_thin_films, magneto_optical_conductivity, PhysRevB.100.195110, PhysRevB.97.081402, observation_excitons_ti, berry_phase_exciton}. Only recently, for instance, the exciton spectrum in Bi$_2$Se$_3$ was shown to exhibit topological properties \cite{topological_excitons}. 
While there are works addressing the role of trivial edge states in the dissociation of excitons in semiconductors \cite{souvik2023, doi:10.1126/science.aal4211, Sui2022, Kinigstein2020, doi:10.1063/1.4968794}, the interaction between bulk excitons in TIs and their topologically protected edge states remains, however, largely unexplored. One recent work studies the interplay between bulk and topological states in the formation of excitons in Bi$_2$Se$_3$ and how these affect the optical response of the TI \cite{bowen2023}.

According to Fermi's golden rule, an exciton is expected to decay elastically into a continuum of states in the presence of a given coupling. Excitons lie within the energy gap and in a trivial insulator there are typically no pure electronic excitations accessible for the exciton to decay into. In general, the usual dissociation channels would be radiative (photons) or non-radiative (phonons) recombination \cite{pelant2012luminescence, segall1968}. Topological insulators, instead, always present edge states connecting the valence and conduction bands,  meaning that in addition to light emission, the exciton can decay into e-h pairs formed by edge states. Generically, one may expect that, when a bulk exciton approaches an edge, it will transform into an edge excitation, eventually decaying into the edge Fermi sea or Tomonaga-Luttinger liquid \cite{Stuhler2020}. We are interested exclusively in this decay channel, so we will disregard photon and phonon emission.

Our principal observation here is that, for sufficiently narrow 2D TI systems (TI ribbons), the electron and hole can also decay onto opposite edges, resulting in a charge separation and eventually in a photovoltaic current.
This is achieved by considering asymmetric ribbons with different edges, where one can induce a preferential decay direction of the charge carriers, thus achieving a charge imbalance or equivalently a chemical potential difference between the edges. Additionally, if the dissociating exciton has non-zero center-of-mass momentum $Q$, then the induced edge charge population will also have finite momentum and velocity, forming a current made of topologically protected carriers.
Tuning the different parameters of the tight-biding model, in our case describing a Bi(111) ribbon, we are able to modify the ratios between the different dissociation channels available, making it possible for the charge separation and the current generation processes to compete with the other recombination mechanisms present.


\section{Results}

\begin{figure}[b]
    \centering
    \includegraphics[width=1\columnwidth]{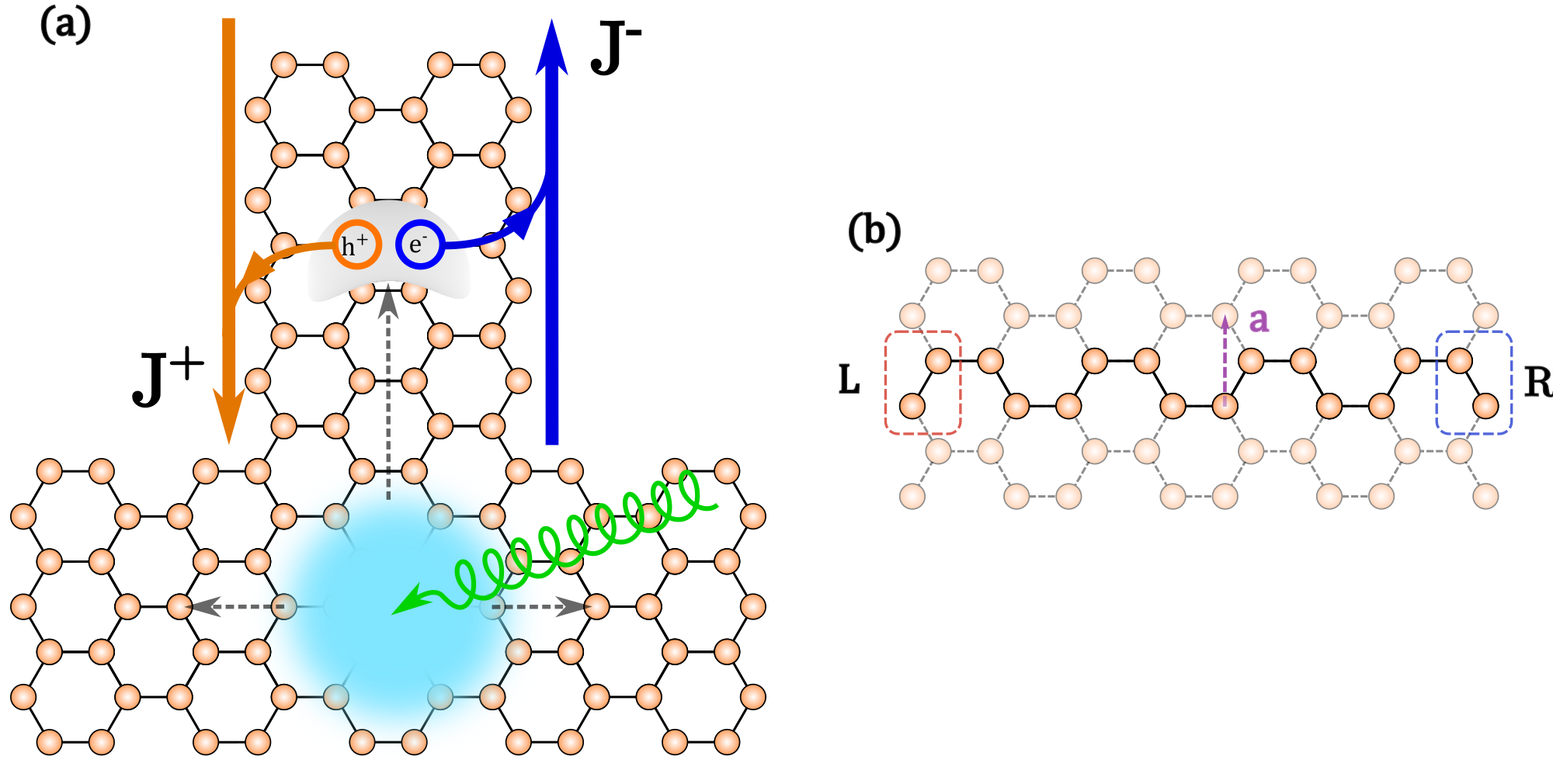}
    \caption{\textbf{Schematic representation of the proposed mechanism}. (a) Device where an exciton wave packet is created at the bulk of the sample, where it will diffuse in any direction. Excitons entering the top ribbon
   present  a finite momentum $Q$, giving rise to an out-of-equilibrium edge carrier population with non-zero momentum and velocity, thus forming a topologically protected current.
    (b) Unit cell of the Bi(111) nanoribbon where the dissociation process takes place. We introduce onsite energies on the (left) edge (atoms labeled as $L$) to split the topological edge bands.}
    \label{fig:setup}
\end{figure}

\begin{figure}[b]
    \centering
    \includegraphics[width=1\columnwidth]{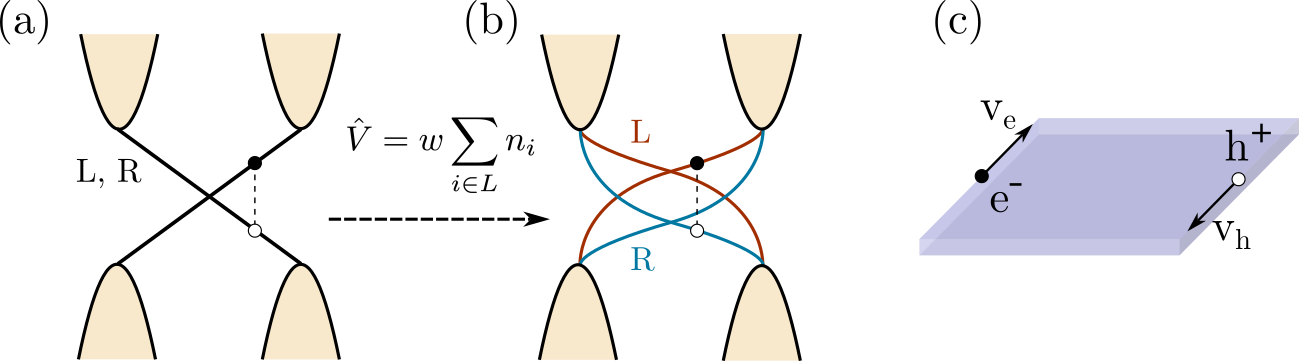}
    \caption{\textbf{Splitting of the edge bands}. (a) For a topological insulator with inversion symmetry, the edge bands of both sides are degenerate, resulting in identical rates for intra-edge and inter-edge transitions. (b) The introduction of an edge offset potential allows to split the edge bands, producing a distinction between the different transitions. (c) For each edge e-h pair we can determine its total velocity as $v_{\rm{e-h}}=v_e-v_h$ to establish whether it carries current or not. For the pair drawn in (b), we observe that $v_e>0$ and $v_h < 0$, meaning that $v_{\rm{e-h}}=v_e-v_h>0$ (see Methods section for the definition of the velocity).}
    \label{fig:edge_bands_splitting}
\end{figure}

A purely electronic exciton decay can take place in the form of a non-interacting e-h pair where both constituents are located on the same edge, or on opposite edges, which may result in charge transportation since edge electrons and holes have typically finite momentum (and velocity). 
Due to time-reversal invariance, however, there is a ${k}\leftrightarrow -{k}$ symmetry in the electronic bands (see Fig. \ref{fig:edge_bands_splitting}a), meaning that the e-h pair can either be equally located at ${k}$ or -${k}$, preventing such possibility for both inter- and intra-edge processes. In short, if a charge current appears in a dissociation process, the total one must be zero since there is always an allowed transition to the time-reversal partner. On top of time-reversal invariance, note also that the system may also possess inversion symmetry, forcing any current appearing on one edge to be cancelled by the one appearing on the opposite one. 

Even if a priori one is unable to generate current in the presence of time-reversal symmetry, it is still possible to generate a charge imbalance between the edges. Consider that we introduce an asymmetry between the edges, via an electric field applied in the direction perpendicular to the infinite edges, or simply by some asymmetric termination.
The latter is implemented in the following Hamiltonian:

\begin{equation}
    H = H_0 + w\sum_{i\in L} n_i,
\end{equation}
where $H_0$ corresponds to a tight-binding model of a ribbon of Bi(111), which is known to be a topological insulator, and the second term is the edge offset potential. In Fig. \ref{fig:setup}(b) we show an example unit cell of the Bi(111) ribbon, and the atoms we identify as left (L) and right (R). On the left ones, we introduce additional onsite energies corresponding to the edge offset, aimed at splitting the edge bands. 
Then, as long as the perturbation does not close the bulk gap, the edge bands will split, as schematically shown in Fig. \ref{fig:edge_bands_splitting}(b). The splitting is expected to produce a different transition rate depending on whether the electron-hole pair is localized on the left-right boundaries, respectively, or the right-left ones. 

In addition to charge accumulation, current generation is also possible if time-reversal symmetry is broken. This may occur by selective population of excitons with non-zero $Q$, avoiding their time-reversal partners with opposite momentum. One possible way to achieve this is shown in Fig. \ref{fig:setup}(a), where an exciton wave packet is created in the bulk of the sample. This packet is generically described by a momentum distribution $\ket{X} = \int d\mathbf{Q}f(\mathbf{Q})\ket{X(\mathbf{Q})}$. Since excitons with finite momentum also have finite velocity, some of them will propagate into the top channel where the edge offset is present. This populates the ribbon with excitons with finite $Q$, but not their time-reversal companions. From the dissociation of these excitons into inter-edge electron-hole pairs we expect to generate a topologically protected photocurrent.

\begin{figure}[h]
    \centering
    \includegraphics[width=\columnwidth]{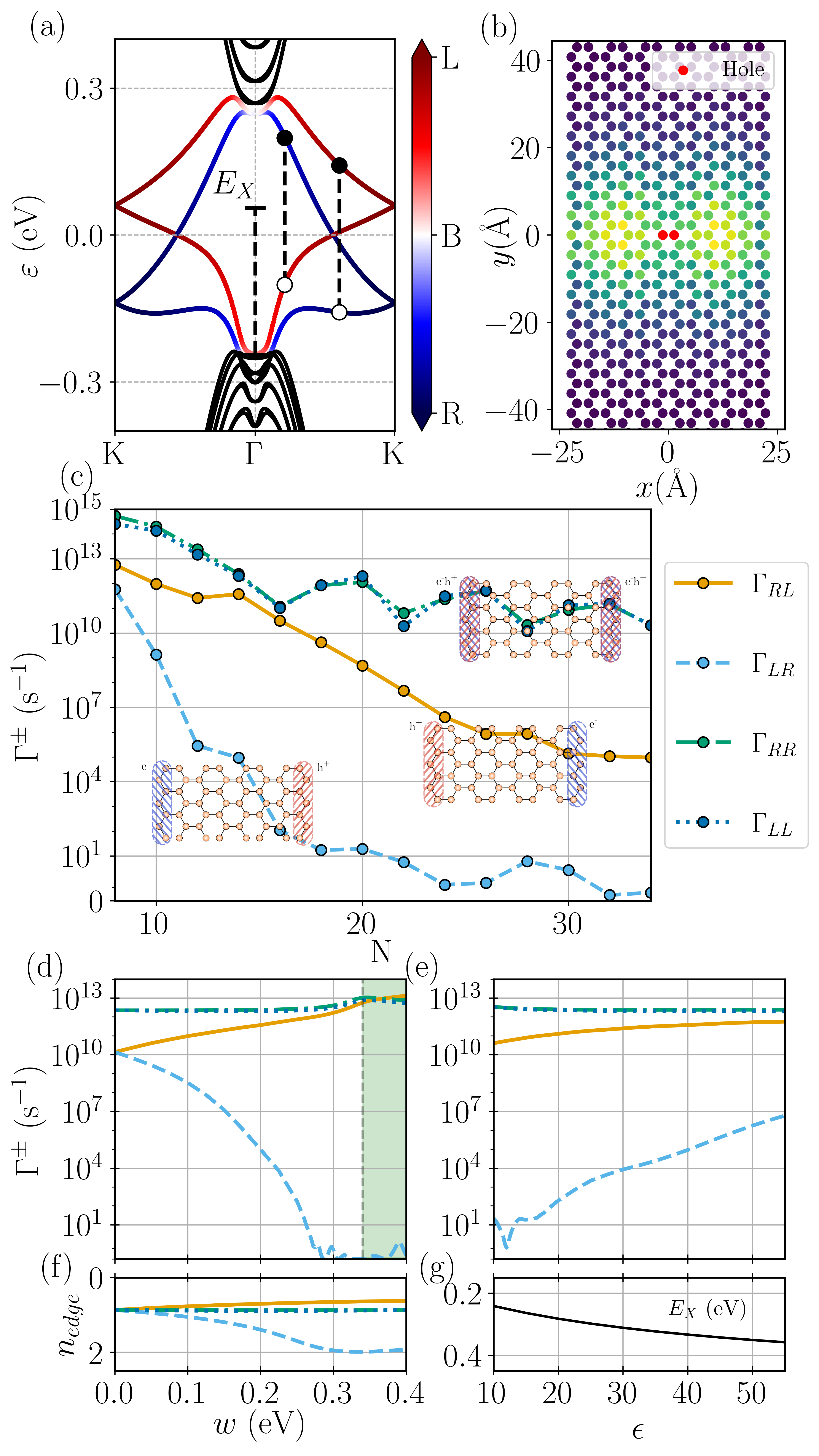}
    \caption{\textbf{Transitions at $Q=0$.} (a) Band structure of the Bi(111) ribbon for $N=20$ and $w=0.2$ eV. The edge bands are colored according to the electronic occupation at the edges of the ribbon. (b) Real-space electronic density probability of the ground state exciton for $N=12$. (c) Transition rates of the ground state exciton to the different edge electron-hole pairs as a function of the width of the ribbon $N$, for $w=0.2$ eV. (d, f) Transition rates and edge occupation as a function of the edge offset potential $w$ for $N=14$. (e, g) Transition rates and ground state exciton energy as a function of the dielectric constant $\varepsilon$ for $N=14$. (c, d, e, f) share the same legend.}
    \label{fig:transitionsQ0}
\end{figure}

To test these hypothesis, we need to evaluate the transition rate from the exciton to each one of the possible electron-hole pairs. Instead of using the band number, we denote each band by its location or edge index, $R$ (right) and $L$ (left). Thus,  for instance, an electron and a hole located on the opposite edges with momentum $k$ would be $\ket{L, R, k}$. With this notation, we want to evaluate the following transition rates:
\begin{equation}
\label{eq:fermi_rule}
    \Gamma^{\pm}_{ss'} = \frac{2\pi}{\hbar}\left|\braket{X|V|s, s', \pm{k}}\right|^2\rho(E_X)
\end{equation}
where $s,s'\in\{R, L\}$ denote the edge where the electron, hole are localized respectively, $\rho$ is the density of states of the final continuum of states, namely the edge e-h pairs, and $V$ is the electrostatic interaction. The initial exciton $\ket{X}$ is taken as the bulk ground state exciton (see Methods section), and $k$ is chosen such that, given $s,s'$, the corresponding e-h pair has the same energy as the exciton. In case of degeneracies the rates are obtained summing over all degenerate states. Here, owing to inversion symmetry, SOC does not lift the band degeneracy resulting in a two-fold degenerate ground state exciton.
 The sign of ${k}$  must also be  specified since there are two possibilities and, in principle, transitions can be asymmetric in $\pm{k}$. 
When the inversion symmetry is removed by the edge offset potential $w$, e.g., at the left boundary, the edge bands, as shown in Fig. \ref{fig:transitionsQ0}(a), are split. We expect now that the inter-edge transition rates $\Gamma_{RL}$ and $\Gamma_{LR}$ will be different as the inter-edge e-h pairs correspond to different $|{k}|$ points (see Fig. \ref{fig:transitionsQ0}(a)), producing a inter-edge charge imbalance in the material. This mechanism would compete with intra-edge transitions $\Gamma_{RR}$ and $\Gamma_{LL}$, where the electron and hole eventually recombine on the same edge. The intra-edge rates serve then as the baseline to estimate the efficiency of the effect.
\\

\textbf{Transition rates for $Q=0$.}

\begin{figure*}[t]
    \centering
    \includegraphics[width=\textwidth]{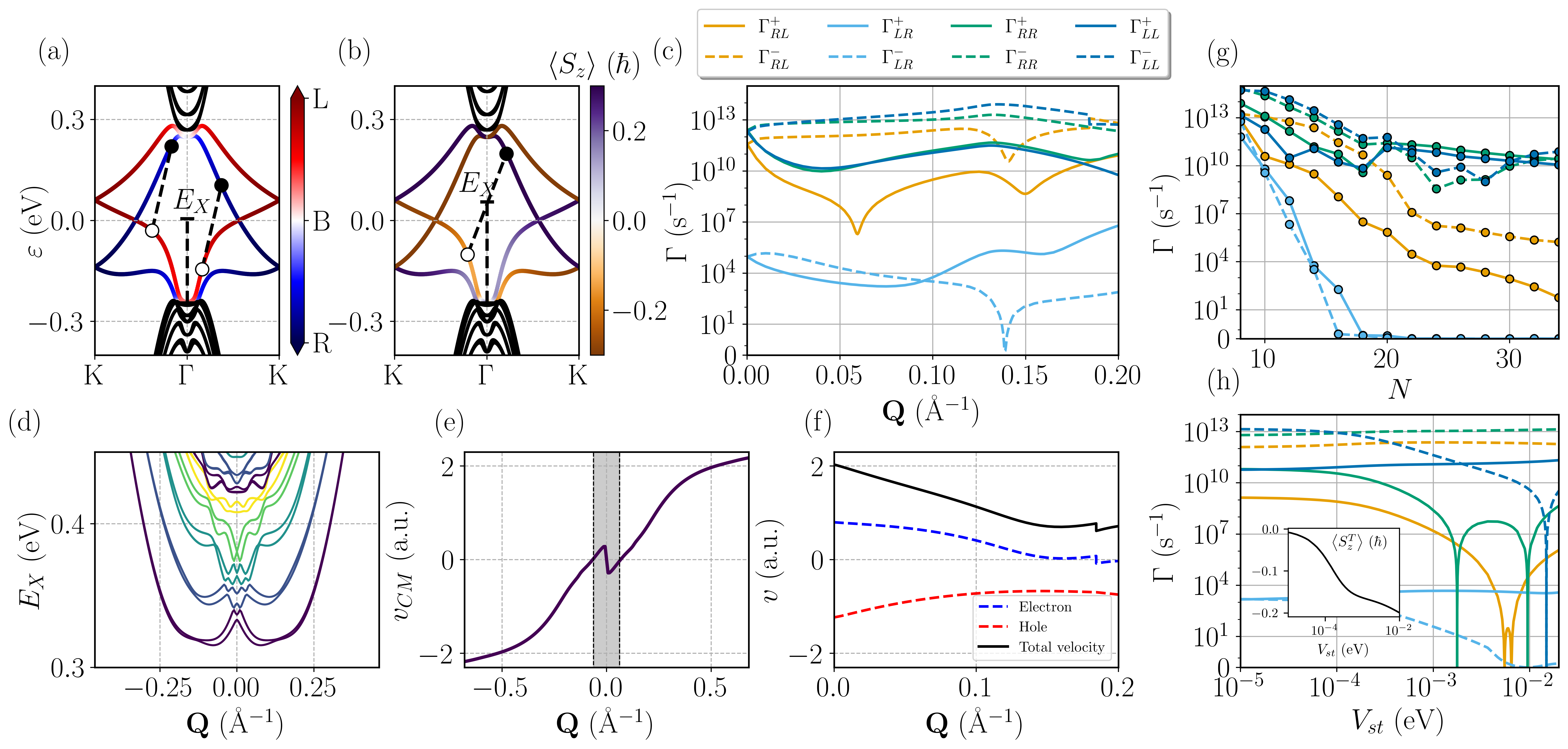}
    \caption{\textbf{Transitions at finite $Q$.} (a, b) Band structure of the Bi(111) ribbon for $N=20$ and $w=0.2$ eV. The first one shows the edge occupation of the bands, while the second ones shows the average spin projection $\left<S_z\right>$ of the bands. (c) Transition rates of the ground state exciton as a function of $Q$ for $N=14$. (d) Low energy exciton band structure. Each color corresponds to four excitonic states in total. (e) Center-of-mass velocity of the ground state exciton. The shadowed region denotes the fraction of excitons that do not contribute to the formation of an edge current. (f) Velocity $v=v_e - v_h$ of the relevant electron-hole pair $\Gamma^-_{RL}$ and of each component individually, for $N=14$, $w=0.2$ eV. (g) Transition rates as a function of $N$, for $Q=0.1$ \AA$^{-1}$ and $w=0.2$ eV. (h) Transition rates as a function of the staggered potential $V_{st}$ for $N=14$ and $w=0.2$ eV. The inset shows the total spin projection of the ground state exciton, $\left<S_z\right>_X$ as a function of the staggered potential.}
    \label{fig:transitionsQ}
\end{figure*}

\noindent 
As we have already noted, because of time-reversal symmetry it can be proven that for $Q=0$ excitons, the transition rates are symmetric in $+{k} \leftrightarrow -{k}$, i.e. $\Gamma^+_{ss'}=\Gamma_{ss'}^-$ (see proof in Supplemental Material).
Although the edge states are helical, bulk excitons are composed of both degenerate bands, which cannot be distinguished by an additional spin quantum number, resulting in their total spin averaging to zero, $\braket{S_z}_{\rm{X}}= 0$. 
The rates in the presence of an onsite potential ($w=0.2$ eV) as a function of the width of the ribbon $N$ are shown in Fig. \ref{fig:transitionsQ0}c. In general, as expected, the inter-edge rates decay faster as a function of $N$ than the intra-edge ones, with  $\Gamma_{RL}$ being several orders of magnitude higher than $\Gamma_{LR}$ for $N\sim 10 - 30$.   
Notably, for intermediate widths ($N\sim 12-16 $), $\Gamma_{RL}^\pm$ turns out to be comparable to intra-edge rates. This can be attributed, in part, to the peculiar real-space electronic probability density of the exciton, which exhibits a p-like character, as shown Fig. \ref{fig:transitionsQ0}(b). Moreover, it is possible to tune the rates to enhance the inter-edge/intra-edge ratio. In Fig. \ref{fig:transitionsQ0}(d) we show the effect of modifying the edge onsite potential $w$. For $w=0$ there is no charge imbalance since both inter-edge rates are equal. As we increase the potential, one rate $\Gamma_{RL}^\pm$ becomes enhanced as it comes closer to the intra-edge rates, while the other $\Gamma_{LR}^\pm$ decreases. The effect of the onsite potential is approximately splitting the edge bands by the same value $w$. Therefore, as we increase $w$, the corresponding edge pairs become increasingly more distant in $|{k}|$. From Fig. \ref{fig:transitionsQ0}(a) we see that those involved in $\Gamma_{LR}^\pm$ get pushed to the high-symmetry point $K$, where the wave functions are fully localized on the edge. On the other hand, for $\Gamma_{RL}^\pm$, the e-h pairs involved get closer to $\Gamma$ ($k=0$), where the functions have a stronger bulk component. Therefore, it is possible to improve the inter-edge/intra-edge ratio by tuning the localization of the e-h pairs on the edge, as seen in Fig. \ref{fig:transitionsQ0}(d). If the edge bands become too far apart, then some of the electron-hole pairs start localizing at different bands, which we indicate with the green region.

A similar discussion can be done with the dielectric constants of the system. We focus on the dielectric constant of the material $\epsilon$, although the same arguments apply to the substrate constant $\epsilon_s$. Tuning  $\epsilon$ produces a change in the exciton energy, which will result in a transition to pairs with different $k$, as shown in Fig. \ref{fig:transitionsQ0}(e,g). In this case, the specific behaviour will be dependent on the form of the bands. Similarly to the onsite potential, changing the exciton energy drastically could result in pairs hosted in a different set of bands from before, although it is not the case for the range of values considered.\\

\textbf{Transition rates for $Q\neq 0$.}

\noindent Next we consider the transition rates for excitons with finite momentum $Q$. As for $Q=0$ excitons, the edge charge accumulation will still be present as long as we keep finite the edge offset term (we again set a fixed value of $w = 0.2$ eV). Now, the main difference with respect to the rates for $Q=0$ excitons comes from the  asymmetry in $k$. Since the initial exciton is not time-reversal invariant (as it has finite momentum $Q$), all the transition rates  $\Gamma^{\pm}_{ss'}$ $\forall s,s'\in\{R,L\}$ will be different (Fig. \ref{fig:transitionsQ}(a) shows schematically the inter-edge processes). Both intra-edge and inter-edge pairs can carry a net current since they have a finite total velocity ${v}_{\rm{e-h}}(k)\neq 0$, but now there will be no exact cancellation between $k$ and $-k$ pairs. 

The results, displayed in Fig. \ref{fig:transitionsQ}(c), show the expected behaviour: as $Q$ becomes non-zero, the $\pm k$ symmetry of the rates is lifted, namely $\Gamma^+_{ss'} \neq \Gamma^-_{ss'} $. We observe that, for the values of $Q$ considered, $\Gamma^{\pm}_{RL}$ and $\Gamma^{\pm}_{LR}$ rates differ by several orders of magnitude, meaning that the charge separation still takes place. We focus our attention again on these inter-edge rates since electron-hole pairs localized on the same edge are assumed not to contribute to the current as they are prone to recombination (in this case via phonon emission first). 
One inter-edge rate ($\Gamma^-_{RL}$) is close in magnitude to the intra-edge ones for all the values of $Q$ considered.
We also see that $\Gamma^-_{RL}$ differs by several orders of magnitude from $\Gamma^+_{RL}$, supporting our hypothesis that an overall edge current can develop in the material since we are inducing an electronic population imbalanced in $k$. For reference we show in Fig. \ref{fig:transitionsQ}(f) the total velocity of the electron-hole pair corresponding to $\Gamma^-_{RL}$, which is non-zero and positive for the values of $Q$ considered. 
Note that the plot only shows values of $Q$ up to 0.2. For higher values of $Q$, the energy of the exciton increases cuadratically (see Fig. \ref{fig:transitionsQ}(d)) and as a consequence there are no longer available edge e-h pairs. Also, as $Q$ increases, it might happen that either the electron or the hole change the band where they are hosted, as illustrated in Figs. \ref{fig:transitionsQ}(a,b). This produces the discontinuity in the rates and the velocities appearing at $Q\approx 0.18$.\\

We still need to determine the probability of the original exciton wave packet to actually enter the ribbon (or the fraction of excitons doing so). To this aim, we compute the total velocity or center-of-mass velocity of the exciton ${v}_{\rm X}$ as a function of $Q$, as shown in Fig. \ref{fig:transitionsQ}(e). Those with ${v}_{\rm X}>0 $ will enter the ribbon. For a small fraction with $Q>0$, highlighted with the gray region, the excitons  have negative velocity, i.e. they move away from the ribbon. For the fraction of excitons of the highlighted region with negative $Q$, they enter the channel, but contribute with opposite currents (due to time-reversal) to the ones with $Q,v_{\rm X}>0$. However, from the exciton bands in Fig. \ref{fig:transitionsQ}(d), we conclude that it is more likely to have a population of excitons satisfying the latter condition, as it corresponds to a lower energy overall. It should be noted that for a conventional semiconductor, the exciton bands would be parabolic meaning that all velocities for positive momentum would also be positive. Thus, this fraction of excitons that hinders the performance of the device is also intrinsic to the topological insulator, but is expected to be small.

As we did for $Q=0$, in Fig. \ref{fig:transitionsQ}(g) we show the behaviour of the transition rates as we increase the width of the ribbon for $Q=0.1$. As expected, the inter-edge rates decay faster than the intra-edge rates. 
Importantly, up to $N=20$, the relevant inter-edge rate $\Gamma^-_{RL}$ is comparable to the intra-edge ones, being the ratio between these rates a rough measure of the efficiency of the system. The opposing rates $\Gamma^{\pm}_{LR}$ become completely suppressed from $N=18$, enhancing the charge separation. As for the ratio between $\Gamma^{+}_{RL}$ and $\Gamma^{-}_{RL}$, it appears to be relatively constant for the widths under consideration.
Finally, we show that it is also possible to engineer the rates by further tuning the spin of the exciton. For $Q=0$ we obtain that $\left<S_z\right>_{\rm{X}}= 0$, due to the bulk bands being degenerate. However, if the exciton had a finite value of the spin, we would expect different values for the rates $\Gamma^{\pm}_{ss'}$ given that the edge bands also present opposite spin when $k \leftrightarrow -k$, as shown in Fig. \ref{fig:transitionsQ}(b). Therefore, if we are able to induce a finite spin in the exciton we might be able to differentiate even more the plus and minus rates $\Gamma^{\pm}$, in addition to the intrinsic asymmetry coming from the finite $Q$. We achieve this introducing a sub-lattice staggered potential that breaks inversion symmetry in the bulk of the material, thereby fully splitting the bulk bands (see Supplemental Material). We show in Fig. \ref{fig:transitionsQ}(h) how for $Q=0.1$ this potential induces a spin in the ground state exciton (inset), and results in the relevant rates $\Gamma^+_{RL}$, $\Gamma^-_{RL}$ deviating even further from each other. Remarkably, some of the intra-edge rates, which may hinder the performance of the device, are strongly suppressed in a wide range of the staggered potential, becoming even zero at particular values.

\section{Conclusions}

We have noted that the edge states of a 2D TI constitute an alternative dissociation path to exciton recombination. To this end, we have fully characterized the exciton spectrum in Bi(111) nanoribbons, and shown that, if one introduces an onsite edge potential to split the edge states, then one can possibly obtain an edge charge imbalance from the dissociation of excitons into non-interacting edge electron-hole pairs. Additionaly, we have shown that, if we are able to generate a population of excitons in the ribbon that is not time-reversal invariant, then an edge current (topologically protected) may develop. Moreover, the corresponding transition rates can be tuned to increase or decrease the strength of the effect. The present arguments are not dependent on the specific shape of the bands, and we expect that they can be applied and tested both theoretically and experimentally with other 2D TIs such as Bi$_4$Br$_4$, which is a room-temperature TI \cite{Shumiya2022}. The foundation of the effect is not exclusive of the dimensionality and can be trivially extended  to three-dimensional TIs.

\section{Methods}

The material of choice here is a monolayer of Bi(111), also known as $beta$-bismuthene, which was predicted \cite{murakami, bi_bl_dft} and observed to be a 2D TI \cite{bi(111)_thinfilms,PhysRevLett.110.176802}. 
We use a Slater-Koster tight-binding model which properly describes the electronic structure of Bi(111) \cite{liu_allen}. Monolayer Bi(111) appears as a puckered honeycomb lattice and presents two possible standard edge terminations: zigzag and armchair \cite{bi_bl_dft}. We consider ribbons with zigzag terminations, as shown in Fig. \ref{fig:setup}(b) or Fig. \ref{fig:transitionsQ0}(b), defining $N$ as the number of Bi dimers across the unit cell. In particular, we choose $N$ to be even so that the space group of the ribbon is symmorphic \cite{symmorphic_ribbon}.


With the material model established, we now turn our attention to the description of excitons. As they are bound electron-hole pairs, one must consider an electrostatic interaction between the electrons of the solid, i.e.:
\begin{equation}
\label{totalHamiltonian}
    H = \sum_{nk}\varepsilon_{nk}c^{\dagger}_{nk}c_{nk} + \frac{1}{2}\sum_{ijkl}V_{ijkl}c^{\dagger}_ic^{\dagger}_jc_lc_k
\end{equation}
$\varepsilon_{nk}$ denotes the bands of the ribbon, and $V_{ijkl}$ are the matrix elements of the electrostatic interaction. The indices $i,j,k,l$ are short-hand notation for pairs of quantum numbers $(n,k)$, with $n$ the band index and $k$ the crystal momentum, which is an scalar since the system is one-dimensional. The exciton states are then given by a superposition of electron-hole pairs between valence and conduction bands:
\begin{equation}
    \ket{X}_{Q} = \sum_{v,c,k}A_{vc}^{Q}(k)c^{\dagger}_{ck+Q}c_{vk}\ket{\Omega}
\end{equation}
where $\ket{\Omega}$ denotes the Fermi sea, which we assume is the ground state of the system, i.e. we work in the Tamm-Dancoff approximation. $Q$ denotes the center-of-mass or total momentum of the exciton. For the bulk excitons we discuss throughout the article, we restrict $v$ and $c$ to bulk bands, i.e. all bands except for the edge ones (those closing the gap).
Then, the problem of determining the exciton coefficients $A_{vc}^{Q}(k)$ amounts to diagonalizing the total Hamiltonian (\ref{totalHamiltonian}) projected over the sector of single electron-hole pairs, i.e. $PHP$. Therefore, we need to obtain the matrix elements of the interacting Hamiltonian represented in the basis of electron-hole pairs. Denoting the electron-hole pairs as $\ket{v,c,k,Q} = c^{\dagger}_{ck+Q}c_{vk}\ket{\Omega}$, the matrix elements $\braket{v,c,k,Q|H|v',c',k',Q}\equiv H_{vv'}^{cc'}(k,k',Q)$ are given by:
\begin{equation}
\begin{split}
    &H_{vv'}^{cc'}(k,k',Q) = \\ &(\varepsilon_{ck+Q} - \varepsilon_{ck})\delta_{cc'}\delta_{vv'}\delta_{kk'} 
    - (D-X)_{vv'}^{cc'}(k,k',Q)
\end{split}
\end{equation}
The $D, X$ terms represent the direct and exchange interaction between the electron and the hole, and are given by some specific matrix elements of the two-body interaction $V$:
\begin{equation}
\label{eq:interaction_matrix_elements}
\begin{split}
    D_{vv'}^{cc'}(k,k', Q) &= V_{ck+Q,v'k',c'k'+Q,vk} \\
     X_{vv'}^{cc'}(k,k', Q) & = V_{ck+Q,v'k',vk,c'k'+Q}
\end{split}
\end{equation}
This approach to excitons has been used previously in several works \cite{macdonald_mos2, hawrylak, PhysRevB.89.235410, nuno_excitons_review}, since it is a simpler and faster alternative to many-body perturbation theory. To compute the excitons, we need to evaluate explicitly the direct and exchange terms (\ref{eq:interaction_matrix_elements}) in terms of the tight-binding Bloch states. If $\{C^{nk}_{i\alpha}\}_{i\alpha}$ are the coefficients corresponding to the eigenstate $\ket{nk}$ in the lattice gauge \cite{10.21468/SciPostPhysCore.6.1.002}, then the interaction matrix elements $D$ can be written as:
\begin{equation}
\begin{split}
    &D_{vc,v'c'}(k, k', Q) = \\
     &\sum_{ij}\sum_{\alpha\beta}(C_{i\alpha}^{ck+Q})^*(C_{j\beta}^{v'k'})^*C_{i\alpha}^{c'k'+Q}C_{j\beta}^{vk}V_{ij}(k'-k)
\end{split}
\label{eq:direct_exchange}
\end{equation}
where
\begin{equation}
    V_{ij}(k'-k) =\frac{1}{N}\sum_{\bold{R}}e^{i(k'-k)R}V(\bold{R} - (\bold{t}_j - \bold{t}_i))
\end{equation}
$\bold{R}$ are Bravais vectors, $R=\norm{\bold{R}}$ and $\bold{t}_i$ are the atomic positions of the motif. $V_{ij}$ denotes a lattice Fourier transform of the potential centered at $\bold{t}_j-\bold{t}_i$.
This expression is obtained evaluating the four-body integral of the interaction in real-space, using the tight-binding approximation (i.e. point-like orbitals) \cite{uria}. It is also possible to derive an alternative expression for the interaction matrix elements with the interaction in reciprocal space through its Fourier series \cite{ridolfi_expeditious, edus}. Given that the platform for our excitons is a semi-infinite ribbon, the real-space approach is better suited since it takes into account the finite boundaries. As for the exchange term $X$, we set $X=0$ assuming that its contribution is negligible.

For the electrostatic interaction, we use the Rytova-Keldysh potential \cite{rytova, keldysh}. This model has gained interest to describe excitons in two-dimensional materials since it includes screening, as opposed to the bare Coulomb interaction. This is particularly relevant since the exact diagonalization approach does not screen the bare interaction, while many-body perturbation theory does, usually through the random-phase approximation \cite{berkeley_gw}. The Rytova-Keldysh potential is given by:
\begin{equation}
    V(\bold{r})=\frac{e^2}{8\epsilon_0\bar{\epsilon} r_0}\left[H_0\left(\frac{\norm{\bold{r}}}{r_0}\right) - Y_0\left(\frac{\norm{\bold{r}}}{r_0} \right)\right]
\end{equation}
where $\bar{\epsilon}$ is an environmental dielectric constant, given by $\bar{\epsilon} = (\epsilon_s + \epsilon_v)/2$, with $\epsilon_s$, $\epsilon_v$ the dielectric constants of the substrate and vacuum respectively. $r_0$ is a screening length, such that $r_0=d\epsilon/(\epsilon_s + \epsilon_v)$, with $d$ the height of the layer and $\epsilon$ the dielectric constant of the material of interest \cite{effective_mass_prada}. 
$H_0$, $Y_0$ are Struve and Bessel functions of second kind respectively. We follow the prescription of \cite{macdonald_mos2} to renormalize the divergence of the potential at $r=0$, by setting $V(0) = V(a)$, $a$ being the lattice parameter of Bi(111). For our monolayer Bi(111), we set the dielectric constant to $\epsilon=40$, and the environmental one to $\bar{\epsilon}=2.45$, corresponding to a $\text{SiO}_2$ substrate.

The edge electron-hole pairs are defined as conventional pairs, but instead we specify the boundary rather than the band number:
\begin{equation}
    \ket{s,s',k} = c^{\dagger}_{sk+Q}c_{s'k}\ket{GS}
\end{equation}
where $c^{\dagger}_{sk+Q}$ creates a conduction electron such that it is located at side $s$ with the specified momentum. The analogue is done with $c_{s'k}$ for the valence hole, with the overall restriction that the energy of the pair is the same as that of the exciton. Note that for complex edge bands, there could be several electron-hole pairs verifying the same conditions, and so all of them should be taken into account in the transition rates.
The calculation of the transition rates is done through eq. (\ref{eq:fermi_rule}), which involves the matrix elements of the electrostatic interaction between the exciton and the final edge electron-hole pair. Expanding the exciton into electron-hole pairs, the transition rates can be cast in terms of the direct and exchange terms of eq. (\ref{eq:interaction_matrix_elements}). Since we neglect the exchange term, the final expression for the rates is:
\begin{equation}
\begin{split}
    &\Gamma^{\pm}_{ss'}=
    \frac{2\pi}{\hbar}\left|\sum_{v,c,k'}A^{Q}_{vc}(k')D^{sc}_{s'v}(k_e,k',Q)\right|^2\rho(E_X)
\end{split}
\end{equation}
The calculation of the velocity is based on the second quantized version of the velocity operator, namely $v = \sum_{ij}v_{ij}c^{\dagger}_ic_j$
where the indices $i,j$ correspond to pairs $(n,k)$. Taking the expected value with the edge electron-hole pair yields the following expression:
\begin{equation}
    v_{e-h}\equiv\braket{v}_{e-h}=v_{ck+Q,ck+Q} - v_{vk,vk}
\end{equation}
this is, the total electronic velocity is given by the velocity of the electron minus the velocity of the hole \cite{kittel}. For the computation of the single-particle velocity elements we refer to \cite{10.21468/SciPostPhysCore.6.1.002}. Anagolously, we can compute the total electronic velocity of the exciton:
\begin{align}
    \nonumber &\braket{v}_X = \\
    \nonumber &\sum_{v,c,k}A^Q_{vc}(k)\left[\sum_{c'}(A_{vc'}^Q(k))^*v_{c'c} - \sum_{v'}(A_{v'c}^Q(k))^*v_{vv'}\right]\\
    &\equiv \braket{v_e}_X - \braket{v_h}_X
\end{align}
Note that as with the electron-hole pair, the velocity of the exciton can be separated into contributions from the electron and the hole separately. In this context, the velocity operator can be rewritten as:
\begin{equation}
    v = \sum_{c,c',k}v_{ck,c'k}c^{\dagger}_{ck}c_{c'k} - 
    \sum_{v,v',k}v_{vk,v'k}c^{\dagger}_{vk}c_{v'k} \equiv v_e - v_h
\end{equation}
where we have defined two new operators. This leads to defining the center-of-mass velocity of the exciton as:
\begin{equation}
    v_X \equiv v_{CM} = \braket{v_e}_X + \braket{v_h}_X
\end{equation}
which is the relevant quantity for determining the propagation direction of the excitons.

All the quantities shown in the article, from the exciton spectrum to the probability densities and the transition rates have been calculated using the Xatu code \cite{uria}. For the calculation of excitons we have used $N_v=N_c=4$, where $N_v$ ($N_c$) is the number of bulk valence (conduction) bands taken into account, and a minimum of $N_k=800$ points in the Brillouin zone to ensure convergence of all transition rates.

\section{Acknowledgement}

The authors acknowledge financial support from Spanish MICINN (Grant Nos. PID2019-109539GB-C43, TED2021-131323B-I00 \& PID2022-141712NB-C21), María de Maeztu Program for Units of Excellence in R\&D (GrantNo.CEX2018-000805-M), Comunidad Autónoma de Madrid through the Nanomag COST-CM Program (GrantNo.S2018/NMT-4321), Generalitat Valenciana through Programa Prometeo (2021/017), Centro de Computación Científica of the Universidad Autónoma de Madrid, and Red Española de Supercomputación.

\bibliography{bib}

\end{document}